\documentclass[twocolumn,showpacs,preprintnumbers,amsmath,amssymb,prl]{revtex4}

\usepackage{graphicx}
\usepackage{dcolumn}
\usepackage{bm}

\begin{document}
\title{An accurate optical lattice clock with $^{87}$Sr atoms}
\author{Rodolphe Le Targat}
\author{Xavier Baillard}
\author{Mathilde Fouch\'e}
\author{Anders Brusch}
\author{Olivier Tcherbakoff}
\author{Giovanni D. Rovera}
\author{Pierre Lemonde}
\email{pierre.lemonde@obspm.fr} \affiliation{
LNE-SYRTE, Observatoire de Paris\\
61, Avenue de l'Observatoire, 75014 Paris, France }

\date{\today}

\begin{abstract}
We report a frequency measurement of the $^1S_0-\,^3P_0$
transition of $^{87}$Sr atoms in an optical lattice clock. The
frequency is determined to be $429\,228\,004\,229\,879\,(5)\,$Hz
with a fractional uncertainty that is comparable to
state-of-the-art optical clocks with neutral atoms in free fall.
Two previous measurements of this transition were found to
disagree by about $2\times 10^{-13}$, {\it i.e.} almost four times
the combined error bar, instilling doubt on the potential of
optical lattice clocks to perform at a high accuracy level. In
perfect agreement with one of these two values, our measurement
essentially dissipates this doubt.
\end{abstract}

\pacs{06.30.Ft,32.80.-t,42.50.Hz,42.62.Fi}
\maketitle

Recent advances in the field of optical frequency metrology make
measurements with a fractional accuracy of $10^{-17}$ or better a
realistic short term goal\,\cite{Hollberg05}. Among other possible
applications ({\it e.g.} a redefinition of the S.I. second,
optical very long baseline interferometry in space, direct mapping
of the earth gravitational using the Einstein effect,...), a very
interesting prospect with measurements at that level is a
reproducible test of Einstein Equivalence Principle by the
repeated determination of the frequency ratio of different atomic
and molecular
transitions\,\cite{Uzan03,Marion03,Bize03,Fischer04,Peik04,Srianand04}.
The topicality of such a test has recently been renewed by
measurements at the cosmological scale which seem to indicate a
slow variation of the electron to proton mass
ratio\,\cite{Reinhold06}. The richness of the test directly
depends on the performance of the clocks that are used but also on
the variety of clock transitions and atomic species on which high
accuracy frequency standards are based.

In that context, optical lattice clocks are expected to play a
central role in the future of this field. They use a large number
of atoms confined in the Lamb-Dicke regime by an optical lattice
in which the first order perturbation of the clock transition
cancels\,\cite{KatoPal03}. Due to the lattice confinement motional
effects, which set a severe limitation to standards with neutral
atoms in free fall\,\cite{Li04,Sterr04}, essentially
vanish\,\cite{Lemonde05}. This gives hope for a ultimate
fractional accuracy better than $10^{-17}$. On the other hand, the
large number of atoms in an optical lattice clock in principle
opens the way to a short term fractional frequency stability
significantly better than $10^{-15}\,\tau^{-1/2}$ with $\tau$ the
averaging time in seconds. In this regime the coherence time of
the laser frequency locked to the clock transition would be
several seconds, possibly tens of seconds\,\footnote{This implies
that the time constant of this servo-loop be shorter than the
coherence time of the free-running probe laser.}. Such a long
coherence time could for instance be used to reduce the width of
the optical resonances in single ion clocks down to or below the
0.1\,Hz range opening new prospects for these devices also.
Finally, the optical lattice clock scheme is in principle
applicable to a large number of atomic species (Sr, Yb, Hg, Ca,
Mg,...) which is a key feature for the fundamental test discussed
above. It is then particularly problematic that the frequency
delivered by the first two evaluated optical lattice clocks, which
both use $^{87}$Sr, disagree by about $2\times
10^{-13}$\,\cite{Takamoto05,Ludlow06}, \emph{i.e.} $3.5$ times the
combined error bar of the measurements.

\begin{center}
\begin{figure}
\begin{center}
\includegraphics[width=0.9\columnwidth]{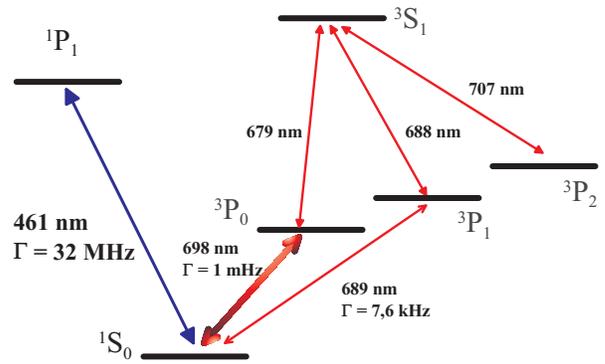}
\end{center}
\caption{\label{fig:levels} Relevant energy levels of $^{87}$Sr.}
\end{figure}
\end{center}

We report here a third independent measurement of the
$^1S_0-\,^3P_0$ transition of $^{87}$Sr in an optical lattice
clock. Our measurement turns out to be in excellent agreement with
the one of Ref.\,\cite{Ludlow06}. This rules out the explanation
of the previous disagreement by an unexpected large frequency
shift which would arise in this new type of clock. This conclusion
is further strengthened by the significantly larger trapping
potential used here as compared to that of both previous
experiments. We operate the optical lattice clock with trapping
depths ranging from $U_0=100\, E_r$ to $U_0=900\, E_r$, where
$E_r$ is the recoil energy associated to the absorption or
emission of a lattice photon ($E_r/h=3.58\,$kHz). We demonstrate
that at such high lattice depths, the residual light shift of the
clock transition is controlled at the Hz level, \emph{i.e.} in the
range of a few $10^{-7}$ of the individual light shift of both
involved atomic states. Including the evaluation of the other
systematic effects, the fractional accuracy of the clock is
$1.2\times 10^{-14}$, comparable to state-of-the-art optical
frequency standards using neutral atoms in free
fall\,\cite{Sterr04,Fischer04}.

\begin{center}
\begin{figure}
\begin{center}
\includegraphics[width=0.9\columnwidth]{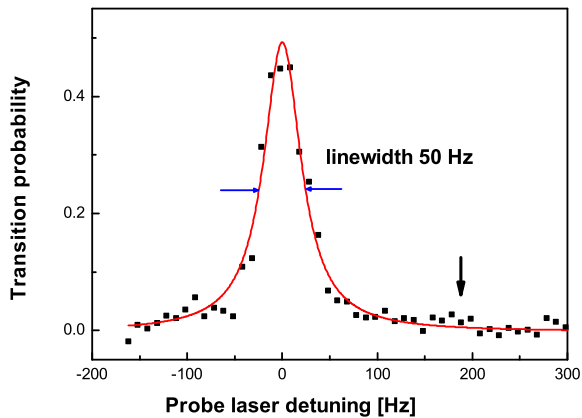}
\end{center}
\caption{\label{fig:resonance} Atomic carrier resonance at 698\,nm
for a probe time of 15\,ms, a power of 3\,$\mu$W and lattice depth
$U_0=170\,E_r$. The vertical arrow is at a detuning corresponding
to the transverse oscillation frequency for this value of $U_0$:
$\nu_{\bot}=186\,$Hz. }
\end{figure}
\end{center}

The Sr optical lattice clock has been described in
Ref.\,\cite{Brusch06} and we only briefly recall the main features
of its operation time sequence. We first load atoms in the
vertical 1-D optical lattice operated at its maximum depth,
presently $150\,\mu$K. The lattice beam is focused to a 90\,$\mu$m
waist and goes through a cloud of cold atoms at 2\,mK that is
produced by a MOT based on the $^1S_0-\,^1P_1$ transition (see the
relevant energy levels in Fig.\,\ref{fig:levels}). Atoms at the
center of the lattice beam are drained to the $^3P_0$ and $^3P_2$
metastable states by a two-photon optical pumping scheme using the
transitions at 688\,nm and 689\,nm. This leads to a continuous
loading of the lattice at a rate of a few $10^4$ atoms/s. After
typically 500\,ms of loading time, the MOT and drain lasers are
switched off. Atoms are repumped back to the $^1S_0$ ground state,
and cooled down to about 10\,$\mu$K in 60\,ms using the narrow
$^1S_0-\,^3P_1$ intercombination transition at 689\,nm. Next, the
lattice depth is ramped down in 1\,ms and the clock transition is
probed by a beam at 698\,nm from an interference-filter stabilized
extended cavity diode laser\,\cite{Baillard06}. Its emission
spectrum is narrowed by Pound-Drever-Hall locking to an
ultrastable cavity\,\cite{Drever83,Quessada03}. The probe beam is
overlapped with the dipole trap and has a waist radius of
210\,$\mu$m. The transition probability to $^3P_0$ is finally
measured by detecting successively the populations of both $^1S_0$
and $^3P_0$ states.

The excitation spectrum at 698\,nm consists in a narrow carrier at
the clock transition frequency that is surrounded by motional
sidebands detuned by the lattice oscillation frequencies. The
carrier is displayed in Fig.\,\ref{fig:resonance}. It has a width
of 50\,Hz, for a probe time of 15\,ms and a laser power of
3\,$\mu$W. The corresponding line Q-factor is about $10^{13}$. We
observed no dependence of the linewidth on the depth of the
trapping potential (the resonance of Fig.\,\ref{fig:resonance} was
for $U_0=170\,E_r$). Depending on the depth of the lattice during
the probe pulse the longitudinal oscillation frequency ranges from
70\,kHz to 210\,kHz, three orders of magnitude largeur than the
width of the resonance. In addition the corresponding sidebands
are reduced by a factor $\eta^2$\,\cite{Leibfried03}, with $\eta$
the Lamb-Dicke parameter ranging from 0.1 to 0.03 here. The
frequency pulling of the carrier by the longitudinal sidebands is
therefore totally negligible. The transverse confinement of the
atoms is much weaker than the longitudinal one and leads to an
oscillation frequency between 140\,Hz and 410\,Hz. Excitation of
the transverse sidebands is possible due to the non-zero
transverse content of the probe wave-vector distribution. This
transverse content results from diffraction (the divergence of the
probe beam is 1\,mrad), the imperfection of the probe/lattice
alignement and wavefront aberrations. However, as can be seen in
Fig.\,\ref{fig:resonance}, transverse sidebands do not emerge from
the noise in the experimental spectrum and are at least 20 times
smaller than the carrier. The corresponding line frequency pulling
is less than 1\,Hz.

\begin{center}
\begin{figure}
\begin{center}
\includegraphics[width=0.9\columnwidth]{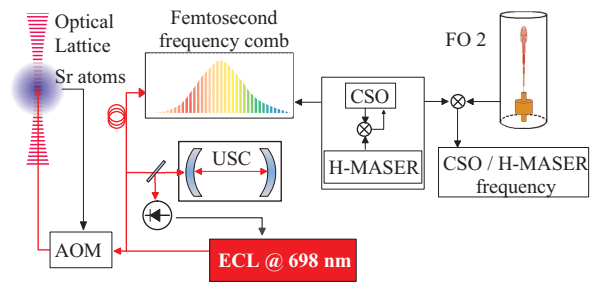}
\end{center}
\caption{\label{fig:setupscheme} Scheme of the frequency
measurement chain : A 698 nm extended cavity laser (ECL) is locked
to an ultrastable cavity (USC) by a Pound Drever Hall servo loop.
The error signal generated by probing the atoms in the optical
lattice is fed-back to control the frequency of the acousto-optic
modulator (AOM). The femtosecond frequency comb is used to measure
the cavity resonance frequency relative to a Cryogenic Sapphire
Oscillator (CSO) locked to a Hydrogen MASER. The measurement of
the frequency of this oscillator by the fountain clock FO2 enables
referencing to the SI second.}
\end{figure}
\end{center}

\begin{center}
\begin{figure}
\begin{center}
\includegraphics[width=0.9\columnwidth]{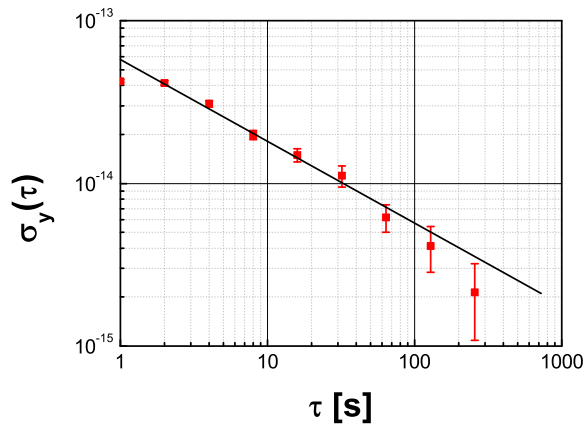}
\end{center}
\caption{\label{fig:variance} Allan deviation of the frequency
measurements. The straight line is a fit by a $\tau^{-1/2}$ law.}
\end{figure}
\end{center}

The frequency measurement chain is depicted in
Fig.\,\ref{fig:setupscheme}. The probe beam frequency is locked to
the atomic transition by alternately probing both sides of the
carrier resonance. The difference between two successive
measurements of the transition probability gives the error signal
that is digitally integrated to servo control the frequency of an
acousto-optic modulator (AOM) used to bridge the detuning between
the atomic transition and the ultra-stable cavity. The cavity
frequency is continuously measured by means of a femtosecond
frequency comb relative to a 1\,GHz
signal\,\cite{Holzwarth00,Jones00}. The latter is provided by an
cryogenic sapphire oscillator\,\cite{Chambon05} that is locked to
a H-maser and referenced to the SI second by the Cesium fountain
clock FO2 with a relative uncertainty in the $10^{-16}$
range\,\cite{Bize04}. The fractional Allan standard deviation of
the measurement is $6\times 10^{-14}\,\tau^{-1/2}$ and is
displayed on Fig.\,\ref{fig:variance}. The frequency resolution is
therefore 1\,Hz (or $2.3\times10^{-15}$ in fractional units) after
about 10 minutes of integration time. The main contributors to the
measurement noise are the femtosecond frequency comb and the
strontium clock with roughly equal weights\,\footnote{The
femtosecond frequency comb measurements average down as
$\tau^{-1/2}$ due to a 200\,ms deadtime between two successive
measurements.}.

\begin{center}
\begin{figure}
\begin{center}
\includegraphics[width=0.9\columnwidth]{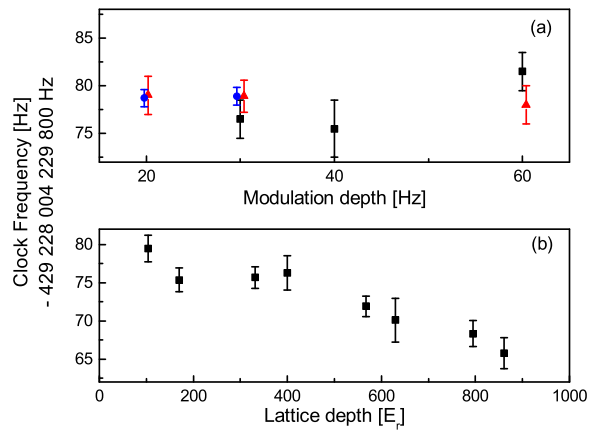}
\end{center}
\caption{\label{fig:latticedepth} (a): Clock frequency as a
function of the depth of the frequency modulation applied for the
frequency lock to the atomic resonance. A modulation depth of
25\,Hz corresponds to locking at half maximum of the resonance.
Measurements are performed for various values of $\alpha$, the
angle formed by the lattice and probe beam polarizations. Both
polarizations are linear. $\blacksquare$: $\alpha=0$,
$\blacktriangle$: $\alpha=\pi/2$, {\Large $\bullet$}:
$\alpha=\pi/4$.(b): Clock frequency as a function of the lattice
depth. For both (a) and (b), the error bars only include the
statistical uncertainty.}
\end{figure}
\end{center}

The residual first order Zeeman effect is the main source of
uncertainty in our measurement. Both clock levels of total
internal angular momentum $F=9/2$ have different Land{\'e} factors
which correspond to a frequency shift of -0.18 Hz/mG/$m_F$ and
-0.08 Hz/mG/$m_F$ for $^1S_0$ and $^3P_0$ respectively, with $m_F$
the magnetic number. The measurements are performed at zero
magnetic field. Three pairs of Helmholtz coils surround the vacuum
chamber and are used to compensate for the environmental field by
minimizing the width of the atomic carrier resonance. The
cancellation is performed to within 10\,mG, limited by
environmental fluctuations. The residual field may result in a
systematic frequency shift if there is an unbalance between the
Zeeman sublevels populations or an asymmetry of the probe laser
polarization with respect to the direction of that field. The
population asymmetry is evaluated by measuring the clock frequency
as a function of the depth of the servo-loop frequency modulation.
The results of this test are shown in
Fig.\,\ref{fig:latticedepth}.(a) for various values of the angle
formed by the probe laser and the lattice beam polarizations
showing no clear dependence of the clock frequency on these
parameters. We conservatively assign an uncertainty of 5\,Hz to
the first order Zeeman effect. This present limitation to our
accuracy budget is purely technical and can certainly be improved
by orders of magnitude by resolving the Zeeman structure and using
spin-polarized atoms.

The lattice wavelength control is presently performed with a
commercial wavemeter which exhibits long term fluctuations of the
order of $10^{-3}$\,nm at 813\,nm. Tuning of the lattice to the
exact point where the first order light shift cancels is therefore
limited to that level and all measurements are performed by
interleaving sequences with different lattice
depths\,\cite{Brusch06}. In Fig.\,\ref{fig:latticedepth}.(b) is
shown the clock frequency as a function of the lattice depth. The
detuning with respect to the magic wavelength deduced from this
particular measurement is $1.1\times 10^{-3}$\,nm. At the average
lattice depth $U_0=400\,E_r$, the differential light shift is
$4.5(9)$\,Hz with an uncertainty which corresponds to $6\times
10^{-7}$ of the individual shift of both clock states. Note that
this evaluation takes into account all contributors to the atomic
dynamic polarizability including the higher order multipolar terms
and the residual vector and tensor terms in the actual
polarization and magnetic state
configuration\,\cite{Ovsiannikov06}. The frequency shift due to
the atomic hyperpolarizability is evaluated by combining the
present measurements with those of Ref.\,\cite{Brusch06}. This
leads to a correction of the measurements by 0.6\,Hz at
$U_0=400\,E_r$ with an uncertainty equal to the correction.

\begin{center}
\begin{figure}
\begin{center}
\includegraphics[width=0.9\columnwidth]{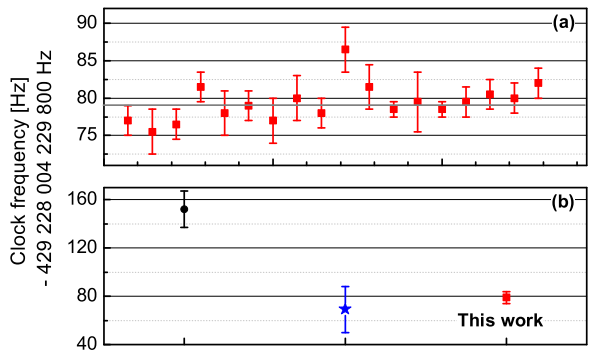}
\end{center}
\caption{\label{fig:frequences} (a): Measurements of the clock
transition with their individual statistical error bar. The
average of these points is $429\,228\,004\,229\,879.4\,$Hz with a
statistical uncertainty of $0.45\,$Hz ($\chi^2=1.0$). The
uncertainty on systematic effects is 5\,Hz. (b): Reported
measurements of the $^1S_0-\,^3P_0$ transition of $^{87}$Sr.
{\Large $\bullet$}: Ref.\,\cite{Takamoto05}, $\bigstar$:
Ref.\,\cite{Ludlow06}, $\blacksquare$: This work.}
\end{figure}
\end{center}

Finally, we have measured the clock frequency as a function of the
atomic density to check for a possible frequency shift due to cold
collisions in our unpolarized fermionic sample. This leads to a
correction of the clock frequency of $1(1)$\,Hz at the average
atomic density of about $10^{11}$at/cm$^{3}$.

\begin{table}
\caption{\label{tab:systematics}Uncertainty budget (all numbers
are in Hz). The shifts due to the lattice are given at
$U_0=400\,E_r$ which is our average lattice depth for this
measurement.}
\begin{ruledtabular}
\begin{tabular}{p{5cm}cc}
 Systematic effects & Correction&Uncertainty\\
\hline
First order Zeeman shift & 0 & 5 \\
Lattice AC Stark shift& 4.5 & 0.9 \\
Lattice 2$^{\rm{nd}}$ order Stark shift\,\footnote{These numbers
result from the data of Ref.\,\cite{Brusch06} combined with new measurements.}  & 0.6  & 0.6 \\
Line pulling by motional sidebands & 0 & $1$\\
Cold collisions & 1 & 1\\
Probe AC Stark shift & 0 & $\ll{1}$ \\
Blackbody radiation shift & 2 & $\ll{1}$\\
FO2 fountain accuracy & 0 & $\ll{1}$\\
\\
\hline
Total & 8 & 5.3 \\
\end{tabular}
\end{ruledtabular}
\end{table}

The accuracy budget of the Sr optical clock is summarized in table
\ref{tab:systematics} and gives a total systematic uncertainty of
5\,Hz. The weighted average of our measurements which are plotted
in Fig.\,\ref{fig:frequences}.(a) is
$429\,228\,004\,229\,879.4$\,Hz with a statistical uncertainty of
0.5\,Hz. In Fig.\,\ref{fig:frequences}.(b) are shown the result of
this new measurement together with the two previously published
values of Ref.\,\cite{Takamoto05} and Ref.\,\cite{Ludlow06}. Our
measurement is in excellent agreement with the one of
Ref.\,\cite{Ludlow06}. This agreement is a confirmation that
optical lattice clocks can indeed operate at a high level of
accuracy.

Today this level is comparable to state-of-the-art optical clocks
with neutral atoms in free fall\,\cite{Sterr04,Fischer04} with
obviously large room for improvement. By polarizing the atoms and
using a small bias magnetic field, we hope to be able to decrease
by orders of magnitude the uncertainty due to the linear Zeeman
effect which is presently the main contributor to the measurement
uncertainty. The evaluation of the other effects will take
advantage of the unprecedented versatility of this new type of
clock. A large number of operation parameters can be varied such
as the trapping depth and configuration (polarization of the
trapping laser, 1D or 3D lattice), the probe polarization, the
atom number, the interrogation scheme (Rabi or Ramsey) and
duration, etc... This will be of great value for untangling the
various physical effects that contribute to the uncertainty
budget.

We thank A. Clairon and S. Bize for useful comments on the
manuscript and together with G. Santarelli, F. Chapelet and M.
Tobar for their work on the FO2 fountain and cryogenic oscillator.
SYRTE is Unit\'e Associ\'ee au CNRS (UMR 8630) and a member of
IFRAF (Institut Francilien de Recherche sur les Atomes Froids).
This work is supported by CNES and DGA.

\bibliographystyle{prsty}

\end{document}